\documentclass[a4paper,10pt,notitlepage,twocolumn,aps,pra,showpacs,showkeys,amsmath,amssymb,nolongbibliography,superscriptaddress,floatfix,nofootinbib,UTF8]{revtex4-1}
\usepackage{mathrsfs,bbm}
\usepackage[colorlinks=true,citecolor=red,filecolor=blue,linkcolor=blue,urlcolor=blue]{hyperref}
\usepackage{geometry}
\begin{document}

\title{Understanding the electromagnetic $4$-potential in the tetrad bundle}

\author{Yufang Hao}
\email[]{haoyufang712@foxmail.com}
\author{Jiayin Shen}
\author{Jianhong Ruan}
\email[Corresponding author:]{jhruan@phy.ecnu.edu.cn}
\affiliation{Department of Physics, East China Normal University, Shanghai 200241, China}

\begin{abstract}
Separation of the spin and orbital angular momenta of the electromagnetic field has been discussed frequently in recent years. The spin and orbital angular momenta cannot be made simultaneously gauge invariant and Lorentz covariant and are not conserved separately. After analyzing the source of the problem, we find that the electromagnetic 4-potential depends on the local reference frame instead of the global reference frame. The transformation of the local reference frame is the intrinsic degree of freedom of the electromagnetic field. Therefore, considering only the Lorentz transformation of the global reference frame and neglecting the Lorentz transformation of the local reference frame may lead to the {noncovariance} of the electromagnetic 4-potential. {Accordingly}, we {redescribe} these difficulties of the electromagnetic field from the perspective of quantum field theory. By using the behavior of the electromagnetic 4-potential that satisfies the Coulomb gauge in Lorentz coordinate transformation, we can construct the electromagnetic vector in the {tetrad bundle}. The various physical quantities {that are induced} by this electromagnetic vector satisfy Lorentz covariance in the {tetrad bundle}. This electromagnetic vector, which is projected onto space-time, is an electromagnetic 4-potential {that satisfies} the Coulomb gauge; thus, the electromagnetic vector is gauge invariant.\\

\noindent {\bf DOI:}\href{https://doi.org/10.1103/PhysRevA.98.033809}{10.1103/PhysRevA.98.033809},\quad{\bf arXiv:}\href{https://arxiv.org/abs/1806.07236}{1806.07236}
\end{abstract}

\maketitle

\section{Introduction}

There has been {substantial} discussion about the separation of {the angular momentum (AM)} of {the} electromagnetic field into its spin and orbital parts \cite{barnett2001optical,bliokh2010angular}. The orbital {AM} density of {the} electromagnetic field is defined as $\mathbf L=E_j\left(\mathbf r\times\nabla\right) \mathscr A_j$ {and the} spin {AM} density {as} $\mathbf S=\mathbf E\times \mathbf{A}$, where $\mathscr A$ is the electromagnetic {$4$-potential and} $\mathbf A$ is the {spatial} part of $\mathscr A$ \cite{darwin1932}. According to this definition, {neither the spin nor the} orbital AM satisfies (\;$U(1)$ group) gauge invariance \cite{barnett2016natures,hammond2017lorentz}. {However,} gauge invariance is an inevitable requirement of an observable physical quantity and the spin and orbital {AMs} of the electromagnetic field can be observed in {various} optical experiments \cite{allen1992orbital,vitullo2017observation}.

This problem can be solved by replacing {$\mathscr A_\mu$} with only its transverse part, which is denoted $A_\mu=\left(0,\mathbf A^{\perp}\right)$ \cite{tannoudji1989atoms,bliokh2013dual,van1994spin}, where $\mathbf A^{\perp}$ satisfies the Coulomb gauge: $\nabla\cdot\mathbf A^{\perp}=0$. Namely, the {definitions} of the spin and orbital {AM} densities {are modified to} $\mathbf L=E_j\left(\mathbf r\times\nabla\right)A_j,\mathbf S=\mathbf E\times \mathbf{A}^{\perp}$. However, this definition {violates} the Lorentz covariance: the Coulomb gauge is not Lorentz covariant. Lorentz covariance is a requirement of the principle of relativity: physical laws should not depend on {the} reference frame.

{The} spin and orbital {AMs cannot} simultaneously be Lorentz covariant and gauge invariant {and} they are not conserved separately \cite{soper2008classical,bliokh2014conservation}. Bliokh {\it et al}. constructed {a set of} conserved spin and orbital {AM densities} \cite{bliokh2014conservation}. {However,} this structure also depends on the Coulomb gauge. {Therefore, it is} not Lorentz covariant. Bliokh {\it et al}. {remark} that this phenomenon is consistent with the experimental operation because a local probe particle will always {identify} a special laboratory reference frame {in which it is at rest}. This explanation is not convincing.
Any observable quantity must be observed and measured in a special laboratory reference frame; {however, most of them do not have a Lorentz-violating mathmatical form because} the Lorentz violation of observation methods {would not} cause {a} Lorentz violation of physical laws. {In other words}, the mathematical form of the measurement result {cannot} depend on the reference frame, which is known as observer Lorentz covariation \cite{bluhm2004breaking}. {Furthermore}, according to the gauge theory of gravitation \cite{utiyama1956invariant}, similarly to [$U(1)$-group] {gauge invariance}, Lorentz covariance is the gauge invariance of {the} $SO(1,3)$ group; {hence, Lorentz covariance} is {also} an inevitable requirement of observable quantities.

However, Bliokh's point of view {provides two main inspirations:}  One is the specificity of the Coulomb gauge.  {Not only can the} Coulomb gauge be used to construct conserved spin and orbital {AMs} but also the canonical quantization procedure {performs well} in this gauge \cite{weinberg1995quantum}.  The other is that the optical phenomenon is  closely related to the reference frame. {Physical laws are local} \cite{weinberg1995quantum,haag2012local}. What if an observable quantity of the electromagnetic field {depends not} on the global reference frame (coordinate system) but {on} the local one (tetrad field)? {We suspect that} the origin of {the noncovariance of the electromagnetic $4$-potential,} which satisfies the Coulomb gauge, is that we {have not} taken the transformation of {the} tetrad field into account. In detail, when the coordinate system is transformed, the choice of tetrad changes, and although {the} Coulomb gauge is broken, the transformation of {the} tetrad will {produce} a phase {that corrects} the deviation [see the Eq. (\ref{pro_trans})].

{According to this view}, we must shift the perspective from space-time to {the tetrad bundle} (see  \cite{kobayashi1963foundations,rudolph2012differential}).  {The} set of all local reference frames of a space-time point $q$  constitutes $q$'s fiber.  As a result, each transformation $\Lambda$ of a {local reference frame} becomes {an intrinsic degree of freedom of the} electromagnetic field. {The} electromagnetic $4$-potential {that} we observed is the projection to space-time of a high-dimensional electromagnetic vectors in {the tetrad bundle}. {This} electromagnetic vector has Lorentz covariance in the high-dimensional {tetrad bundle}. {The main} essence of  {Lorentz covariance is} that the mathematical form of the physical quantity {cannot depend on a reference frame; hence, the} physical quantity should be a ``geometric invariant''. Not all geometric invariants must be vectors {(or tensors)} in space-time; {however,} we {used to replace ``Lorentz covariance'' with ``Lorentz covariance of space-time vectors (tensors)'' narrowly}. The electromagnetic vector in {the tetrad bundle} that we construct is such an example; it is a geometric invariant but {does not have Lorentz covariance when it is projected to space-time,} where we {cannot} obtain all its information.

The {remainder of the} paper is organized as follows: In Secs. \ref{2} and \ref{3}, we restate the issue about Lorentz covariance and gauge invariance {that relates} to the  {AM} of {the} electromagnetic field from the perspective of quantum field theory, which lays the groundwork for Sec.\ref{4}. In Sec.\ref{4},  we {present} the revised {definitions} of various physical {quantities} of {the} electromagnetic field in {the tetrad bundle} and the relationship with {the corresponding classical definition} in space-time. {Section \ref{5} provides a summary}. {Throughout the text}, we use Einstein's sum rule; natural electrodynamical units, {namely,} $\mu_0=\varepsilon_0=c=1$; the Minkowski metric, {which is expressed as} $\eta_{\mu\nu}=diag\left(-1,1,1,1\right)$; Greek indices $\rho,\mu,\nu\cdots=0,1,2,3$; {and} Latin indices $i,j,k\cdots=1,2,3$. We do not distinguish {between} the {notions of a ``vector''} in tangent space and that of a ``1-form'' in the cotangent space; both are called vectors.

\section{energy-momentum tensor}\label{2}
We assume that $\phi^\rho$ is a spin-$1$ vector field with a mass under the Lorentz transformation of the reference frame $x\to\Lambda x$, which is transformed as a vector representation of the Lorentz group \cite{srednicki2007quantum}:
\begin{equation}\label{vector-trans}
U(\Lambda)\phi^\rho(x)U^{-1}(\Lambda)={\Lambda_\sigma}^\rho\phi^\sigma(\Lambda x).
\end{equation}
 {Hence,} $\phi^\rho$ is a $4$-vector.   The canonical energy-momentum tensor of $\phi^\rho$, {which is denoted as $T_N$}, is usually defined as
 \begin{equation}\label{zhengzeT}
T_N^{\mu\nu}=\eta^{\mu\nu}\mathscr L-\frac{\partial\mathscr L}{\partial\left(\partial_\mu \phi^\rho\right)}\partial^\nu \phi^\rho,
\end{equation}
where $\mathscr L$ is the Lagrangian density. 

Using Eq. (\ref{vector-trans}), we can prove that under the coordinate transformation $x\to\Lambda x$,   
\begin{equation}\label{TLorentInva}
U(\Lambda)T_N^{\mu\nu}(x)U^{-1}(\Lambda)={\Lambda_\rho}^\mu{\Lambda_\sigma}^\nu T_N^{\rho\sigma}(\Lambda x)
\end{equation}
Therefore, $T_N$ is a Lorentz tensor with Lorentz covariance. However, $T_N$ {does not have local}  $U(1)$ gauge invariance {because} under the gauge transformation $\phi^\rho(x)\to e^{\mathrm i\epsilon(x)}\phi^\rho(x)$, $\partial^\nu\phi^\rho$ will produce an additional factor, {namely,} $\partial^\nu\epsilon(x)$, which {cannot} be canceled {unless} $\partial_\mu$ {is replaced} with covariant derivative $D_\mu$ in {the} $\mathfrak u(1)$ algebra \cite{zeidler2011quantum}.

Using {the} canonical energy-momentum tensor, we can construct a Noether flow, {namely, the} Lorentz generator density: 
\begin{equation}
M_N^{\rho\mu\nu}=x^\mu T_N^{\rho\nu}-x^\nu T_N^{\rho\mu}.
\end{equation}
By integrating this Noether flow (volume integrals for sufficiently localized fields are assumed), we can {obtain} the generator of the Lorentz group,
\begin{equation}
L^{\mu\nu}=\int M_N^{0\mu\nu}\,\mathrm d^3x,
\end{equation}
where {the spatial part, namely, $L^{ij}$ of $L^{\mu\nu}$,} is the {AM} generator.  However, the canonical energy-momentum tensor {does not} satisfy {index} symmetry generally, {i.e.,} $T_N^{\mu\nu}\not=T_N^{\nu\mu}$. {Index} symmetry is a necessary and sufficient condition for the conservation of $M^{\rho\mu\nu}$ {because}
\begin{equation}\label{shouheng}
\partial_\rho M_N^{\rho\mu\nu}=2T_N^{[\mu\nu]},
\end{equation}
where $[\cdot]$ represents the tensor's anticommutator.  Therefore, the Belinfante energy-momentum tensor was introduced by adding an intrinsic {spin} term to the canonical energy-momentum tensor \cite{belinfante1940current},
\begin{equation}
T_B^{\mu\nu}=T_N^{\mu\nu}+\frac{1}{2}\partial_\rho\left(S^{\rho\mu\nu}-S^{\mu\rho\nu}-S^{\nu\rho\mu}\right),
\end{equation}
where the spin term, {which is denoted as} $\partial_\rho S^{\rho\mu\nu}$, cancels out the antisymmetric part of $T_N$, {thereby} leaving only the {symmetric} part:  
\begin{equation}\label{STT}
\frac{1}{2}\partial_\rho S^{\rho\mu\nu}=-T_N^{[\mu\nu]}.
\end{equation}
Thus, the Belinfante energy-momentum tensor is a symmetric tensor {and} the Lorentz-group generator density, which is denoted by $M_B$, which is induced by the Belinfante energy-momentum tensor, satisfies
\begin{equation}
\begin{split}
M_B^{\rho\mu\nu}&=x^\mu T_B^{\rho\nu}-x^\nu T_B^{\rho\mu}
\\
&=M_N^{\rho\mu\nu}+S^{\rho\mu\nu}+\partial_\kappa\Psi^{\kappa\rho\mu\nu},
\end{split}
\end{equation}
where $\Psi^{\kappa\rho\mu\nu}$ is a surface term {that consists} of coordinates and spins.  We interpret $M_N^{0ij}$ as the orbital {AM} density {and} $S^{0ij}$ as the spin {AM} density.  {The} total Noether flow is conserved:
\begin{equation}\label{Mshouheng}
\partial_\rho M_B^{\rho\mu\nu}=0.
\end{equation}

The famous physicist {S. Weinberg presented} the expression of spin $S^{\rho\mu\nu}$ in Ref.  \cite{weinberg1995quantum}:
\begin{equation}\label{xianxingS}
S^{\rho\mu\nu}=\frac{\partial\mathscr L}{\partial\left(\partial_\rho \phi_\nu\right)}\phi^\mu-\frac{\partial\mathscr L}{\partial\left(\partial_\rho \phi_\mu\right)}\phi^\nu.
\end{equation}
{The} Lorentz covariance condition of the spin {is}
\begin{equation}\label{SLorentInva}
U(\Lambda)S^{\rho\mu\nu}U^{-1}(\Lambda)={\Lambda_\gamma}^\rho{\Lambda_\alpha}^\mu{\Lambda_\beta}^\nu S^{\gamma\alpha\beta}(\Lambda x).
\end{equation}

\section{Restatement of the angular momentum problems}\label{3}
Problems arise in the construction of a massless vector field $A_\mu(x)$ {that is} modeled on a mass vector field $\phi^\rho$ \cite{weinberg1995quantum}. Simply using the creation and annihilation operators of the photon, which are denoted $a^\dagger$ and $a$,  we {cannot} construct a $4$-vector that satisfies (\ref{vector-trans}); $A_\mu(x)$ can only have the form
\begin{equation}
\begin{split}
&A_\mu(x)=(2\pi)^{-\frac{3}{2}}\sum_{h=\pm 1}\int \frac{\mathrm d^3p}{\sqrt{2p^0}}\,\bigg(e_\mu(\mathbf p,h)\\
&\times e^{\mathrm ip_\mu x^\mu}a(\mathbf p,h)
+e_\mu^*(\mathbf p,h)e^{-\mathrm ip_\mu x^\mu}a^\dagger(\mathbf p,h)\bigg),
\end{split}
\end{equation}
where $\mathbf p$ is {the} $3$-momentum, $p=\left(p^0,\mathbf p\right)$ is a {light-like} $4$-momentum, and $h$ is the helicity of the photon. The coefficient $e_\mu(\mathbf p,h)={R(\hat{\mathbf p})_\mu}^\nu e_\nu(\mathbf k,h)$, where  $R(\hat{\mathbf p})$  is a {rotational} transformation that rotates the {spatial} part $\mathbf k=(0,0,1)$ of the standard momentum $k=(1,0,0,1)$ to the direction of $\mathbf p$, which is written as $\hat{\mathbf p}$, and $e_\nu(\mathbf k,h)$ can be expressed as
\begin{equation}
e_\nu(\mathbf k,h)=\frac{1}{\sqrt 2}(0,1,\mathrm ih,0).
\end{equation}
In this configuration, $A_\mu$ satisfies the Coulomb gauge (in vacuum):
\begin{equation}\label{kulun}
A_0=0,\;\partial^j A_j=0.
\end{equation}
Under a {reference-frame} transformation $x\to\Lambda x$, $A_\mu$ behaves as {follows:} 
\begin{equation}\label{transformA}
U(\Lambda)A_\rho(x)U^{-1}(\Lambda)={\Lambda^\sigma}_\rho A_\sigma(\Lambda x)+{\Lambda^\sigma}_\rho\left(\partial_\sigma\Omega\right)(\Lambda x,\Lambda).
\end{equation}
{The} transformation of {the} reference frame leads $A_\mu$  to produce a gauge $\partial_\rho\Omega_\Lambda$ that depends on the reference frame, {where} $\Omega(x,\Lambda)$ can be expressed as a linear combination of creation and annihilation operators. {Reference}  \cite{weinberg1995quantum} {did not present an} explicit expression {for} $\Omega$. According to our calculations, 
\begin{equation}\label{O}
\begin{split}	
\Omega(x,\Lambda)
&=(2\pi)^{-\frac{3}{2}}\sum_{h=\pm 1}\int \frac{\mathrm d^3 p}{2\sqrt{p^0}}\,\bigg(\big[\alpha(p,\Lambda)\\
&+\mathrm ih\beta(p,\Lambda)\big]e^{\mathrm ip_\mu x^\mu}a(\mathbf p,h)+\big[\alpha(p,\Lambda)\\
&-\mathrm ih\beta(p,\Lambda)\big]e^{-\mathrm ip_\mu x^\mu}a^\dagger(\mathbf p,h)\bigg),
\end{split}
\end{equation}
where $\alpha$ and $\beta$ are parameters that depend on the Lorentz transformation $\Lambda$ and the $4$-momentum $p$.  The little group representation of $\Lambda$ is expressed as $W=L^{-1}(\Lambda p)\Lambda L(p)$, {where} $L(p)$ is the standard Lorentz transformation, {which is applied} to boost the standard momentum $k^\mu$ to the $4$-momentum $p^\mu$, {that is}, $L(p)k=p$.
The definitions of $\alpha$ and $\beta$ are $\alpha={W^0}_1$ and $\beta={W^0}_2$. 

From the above discussion, {although both electromagnetic $4$-potential $A_\mu$ and Lorentz vector $\phi^\rho$ have the same index}, their transformation properties are different. {According to} gauge field theory  \cite{zeidler2011quantum}, $A_\mu$ is a gauge potential in the principal bundle whose structure group is $U(1)$, {whereas} $\phi^\rho$ is a vector (component) in the representation space.  Therefore, if we directly apply the definitions of energy-momentum tensor, spin, Lorentz generator density, etc., of $\phi^\rho$ to the electromagnetic $4$-potential $A_\mu$, {problems} will inevitably {be encountered}. {However, researchers typically apply} these definitions directly to electromagnetic fields.  We believe that this is why the orbital and spin {AMs} of the electromagnetic field do not have Lorentz covariance or gauge invariance.

The Lagrangian and action of {the} electromagnetic field in {a} vacuum ({with} no matter) can be expressed as
\begin{equation}
\begin{split}
&\mathscr L(x)=-\frac{1}{4}F_{\mu\nu}(x)F^{\mu\nu}(x),
\\
&S_\gamma=-\frac{1}{4}\int\mathrm d^4x\,F_{\mu\nu}F^{\mu\nu},
	\end{split}
\end{equation}
where ${F_{\mu\nu}(x)=\partial_\mu\mathscr A_\nu(x)-\partial_\nu \mathscr A_\mu(x)}$ is the electromagnetic field tensor {and} $\mathscr A_\mu(x)=A_\mu(x)+\partial_\mu\theta(x)$ is an electromagnetic $4$-potential under any gauge.   We temporarily imitate Eq. (\ref{zhengzeT}) to calculate the canonical energy-momentum tensor of the electromagnetic field,
\begin{equation}\label{TN}
\begin{split}
T_N^{\mu\nu}&=\eta^{\mu\nu}\mathscr L-\frac{\partial\mathscr L}{\partial\left(\partial_\mu\mathscr A_\rho\right)}\partial^\nu \mathscr A_\rho\\
&=-\frac{1}{4}\eta^{\mu\nu}F_{\rho\sigma}F^{\rho\sigma}+F^{\mu\rho}\partial^\nu\mathscr A_\rho,
\end{split}
\end{equation}
and imitate Eq. (\ref{xianxingS}) to calculate the spin term $S^{\rho\mu\nu}$ of the electromagnetic field:
\begin{equation}\label{S}
\begin{split}
S^{\rho\mu\nu}&=\frac{\partial\mathscr L}{\partial\left(\partial_\rho\mathscr A_\nu\right)}\mathscr A^\mu-\frac{\partial\mathscr L}{\partial\left(\partial_\rho \mathscr A_\mu\right)}\mathscr A^\nu\\
&=F^{\rho\mu} \mathscr A^\nu-F^{\rho\nu}\mathscr A^\mu.
\end{split}
\end{equation}

According to Eqs. (\ref{TN}) and (\ref{S}), it is easy to calculate the three-dimensional form of {the AM} density; {the calculation is typically presented} in textbooks \cite{jackson1999classical},
\begin{equation}\label{jiaodongliang}
\mathbf L=E_j\left(\mathbf r\times\nabla\right) \mathscr A_j
,\quad
\mathbf S=\mathbf E\times\mathbf A,
\end{equation}
where $\mathscr A_\mu=(\varphi,\mathbf A)$.

There {is no} local gauge invariance in Eqs.(\ref{TN}) and (\ref{S}). Unlike the gauge transformation of a vector field $\phi^\rho$, the gauge transformation of $\mathscr A_\mu(x)$ is not {performed by multiplying} a local phase factor $e^{\mathrm i\epsilon(x)}$. $\mathscr A_\mu(x)$ is the gauge potential in {the} $\mathfrak u(1)$ algebra, i.e. \cite{zeidler2011quantum},
\begin{equation}\label{gauge-trans}
\mathscr A'_\rho(x)=\mathscr A_\rho(x)+\partial_\rho\epsilon(x).
\end{equation}
Substituting (\ref{gauge-trans}) into (\ref{TN}) and (\ref{S}) will yield {two} additional terms: $F^{\mu\rho}\partial^\nu\partial_\rho\epsilon$ and $2F^{\rho[\mu}\partial^{\nu]}\epsilon$. {The} canonical energy-momentum tensor and spin {do not satisfy} $U(1)$ gauge invariance. We {cannot} change this {phenomenon} even if we replace the derivative operator $\partial_\mu$ in Eq. (\ref{TN}) with the covariant derivative $D_\mu$.  

The {typical} solution is to replace $\mathscr A_\rho$ by its  transverse part, {which is denoted as} $A_\rho$ \cite{tannoudji1989atoms,bliokh2013dual,van1994spin} {and always} satisfies {the} Coulomb gauge; {hence, it is invariant} under {the} gauge transformation {in} Eq. (\ref{gauge-trans}):
\begin{align}
\label{gau-invTN}
T_N^{\mu\nu}&=-\frac{1}{4}\eta^{\mu\nu}F_{\rho\sigma}F^{\rho\sigma}+F^{\mu\rho}\partial^\nu A_\rho,
\\
\label{gau-invS}
S^{\rho\mu\nu}&=F^{\rho\mu}A^\nu-F^{\rho\nu}A^\mu,
\end{align}
The three-dimensional form of {the} orbital and spin {AMs} is corrected {by}
\begin{equation}\label{gauge-invLS}
\mathbf L=E_j\left(\mathbf r\times\nabla\right)A_j
,\quad
\mathbf S=\mathbf E\times\mathbf A^{\perp},
\end{equation}
where $A_\mu=(0,\mathbf A^\perp)$ and $\nabla\cdot\mathbf A^\perp=0$.

However, the frame transformation, (\ref{transformA}), of $A_\rho$ {causes} a new problem immediately: $T_N^{\mu\nu}$ and $S^{\rho\mu\nu}$ no longer satisfy {the} Lorentz covariance conditions in Eqs. (\ref{TLorentInva}) and (\ref{SLorentInva}) and neither do the orbital and spin {AM} densities {that are} induced by them. The source of the {difficulty} is the dependency of {the} Coulomb gauge on {the} reference frame: if we {perform} a boost transformation to the reference frame $\bar x=\Lambda x$ {while transforming} $A_\mu(x)$ {via the classic approach, namely,} $\bar A_\mu(\bar x)={\Lambda_\mu}^\rho A_\rho(\Lambda^{-1}\bar x)$, then $\bar A_\mu(\bar x)$ no longer satisfies this gauge. {However,} the Belinfante {energy-momentum} tensor that sums the contributions of {the} orbital and spin {parts} (ignoring the surface terms), {which is expressed as}
\begin{equation}
T_B^{\mu\nu}=-\frac{1}{4}\eta^{\mu\nu} F_{\rho\sigma}F^{\rho\sigma}+F^{\mu\rho}{F^\nu}_\rho,
\end{equation}
is {index symmetric}, gauge invariant, and Lorentz covariant. {Hence,} it is a {well-defined} observable quantity.  

The origin of these difficulties in the final analysis is that $A_\mu(x)$'s transformation, (\ref{transformA}), differs substantially from the $4$-vector $\phi^\rho$. We {cannot} simply assert that $A_\mu(x)$ {does not} have Lorentz covariance.  After {careful consideration}, we find that the essential requirement of Lorentz covariance is that the definition of a physical quantity be independent of {the} reference frames, that is, {the} physical quantity is a {geometric invariant}. The geometric invariants of different structures have different Lorentz transformation forms. {The} most familiar example, {namely,} the Christoffel symbol, {which is denoted as} ${\Gamma^\rho}_{\nu\mu}$ is a projection to space-time of the torsion-vanishing metric connection $\omega$ in the {tetrad bundle}. This projection mapping {depends on the} reference frames, {whereas} the mathematical definition of $\omega$ is independent of {the} reference frames \cite{kobayashi1963foundations}.  As a projection of a geometric invariant, the Christoffel symbol's transformation {differs substantially} from that of a tensor. We {cannot simply treat} the ``Lorentz transformation'' as the ``Lorentz transformation of tensors.''

If we can {identify} a geometric invariant whose projection to space-time is the electromagnetic $4$-potential, then we will prove that the electromagnetic $4$-potential {remains} Lorentz covariant {and} the problem will be solved. Starting from the Lorentz transformation {in} Eq. (\ref{transformA}) of $A_\mu(x)$, we modify the definitions of various physical quantities of the electromagnetic field.

\section{Promotion of the Electromagnetic $4$-Potential to a Bundle Vector}\label{4}
{According to Eq. (\ref{transformA}) and the little group representation of the unit Lorentz transformation $\mathbbm 1$, which is expressed as $W(p,\mathbbm 1)=L^{-1}(\mathbbm 1p)\mathbbm 1L(p)=\mathbbm 1$,} $\alpha(p,\mathbbm 1)=0,\;\beta(p,\mathbbm 1)=0$, which leads to $\Omega(x,\mathbbm 1)=0$. 

{We rewrite} Eq. (\ref{transformA}) as
\begin{equation}\label{A+O}
\begin{split}
	U(\Lambda)&\big(A_\rho(x)+\partial_\rho\Omega(x,\mathbbm 1)\big)U^{-1}(\Lambda)
	\\&={\Lambda^\sigma}_\rho \big(A_\sigma(\Lambda x)+\left(\partial_\sigma\Omega\right)(\Lambda x,\Lambda)\big)
\end{split}
\end{equation}
{and observe that} $A_\rho+\partial_\rho\Omega$ behaves {similar to} a $4$-vector under a {reference-frame} transformation.  We define 
\begin{equation}\label{invA}
	\mathcal A_\rho(x,\Lambda)=A_\rho(x)+\partial_\rho\Omega(x,\Lambda).
\end{equation}
{Then,} according to Eq. (\ref{A+O}), {the} Lorentz transformation of $\mathcal A_\rho$ and its derivative can be {obtained directly:}
\begin{equation}
\begin{split}
	&U(\Lambda)\mathcal A_\rho(x,\mathbbm 1) U^{-1}(\Lambda)={\Lambda^\sigma}_\rho \mathcal A_\sigma(\Lambda x,\Lambda),
	\\
	&U(\Lambda)\partial_\sigma\mathcal A_\rho(x,\mathbbm 1) U^{-1}(\Lambda)={\Lambda^\nu}_\sigma{\Lambda^\mu}_\rho\left(\partial_\nu\mathcal A_\mu\right)(\Lambda x,\Lambda).
\end{split}
\end{equation}	
Although $\mathcal A_\rho$ is similar to a $4$-vector, {$\mathcal A_\rho$} depends on both $x$ and $\Lambda$. {Hence,} $\mathcal A_\rho$ is not a vector field in space-time.   If we {consider} $(x^\rho,{\Lambda^\mu}_\nu)$ as a coordinate, {$x$ and $\Lambda$} will constitute a coordinate domain of the {tetrad bundle}.  For a point $(q,e_0,e_1,e_2,e_3)$ in the {tetrad bundle}, {where} $q$ is a point in space-time whose coordinate is $x^\rho$, {any orthonormal basis} $(e_0,e_1,e_2,e_3)$ of the tangent space of $q$ can be expressed as $e_\nu={\Lambda^\mu}_\nu\partial_\mu$. {Hence,} ${\Lambda^\mu}_\nu$ is {selected} as the coordinate of this basis which is called a tetrad of $q$. Therefore, $\mathcal A_\rho$ is a vector field in the {tetrad bundle}. {Strictly}, $\mathcal A_\rho$ is a component of the vector field and $\mathcal A=\mathcal A_\rho\mathrm\,\mathrm dx^\rho$ is a vector field {where $\mathrm dx^\rho$ is a vector not in space-time but in the tetrad bundle}. Both $\mathrm dx^\mu$ and $\mathrm d{\Lambda^\rho}_\sigma$ constitute the vector {basis} of {the} cotangent space of $(q,e_0,e_1,e_2,e_3)$. We rewrite Eq. (\ref{invA}) {in} a {coordinate-independent} form and $(q,e_0,e_1,e_2,e_3)$ can be simplified as $(q,e)$:
\begin{equation}\label{bundleform}
\mathcal A(q,e)=\mathcal A_\rho(q,e)\,\mathrm dx^\rho+{\mathcal A^\nu}_\mu(q,e)\,\mathrm d{\Lambda^\mu}_\nu,
\end{equation}
{where}
\begin{align}
	\label{Aqe}
	&\mathcal A_\rho(q,e)=A_\rho(x)+\partial_\rho\Omega(x,\Lambda),
	\\
	&{\mathcal A^\nu}_\mu(q,e)=0.
\end{align}
{
Although it vanishes, ${\mathcal A^\nu}_\mu$ is written in Eq. \eqref{bundleform} to emphasize the difference between the electromagnetic vector $\mathcal A(q,e)=\mathcal A_\rho(q,e)\,\mathrm dx^\rho$ and electromagnetic 4-potential $A(q)=A_\rho(x)\,\mathrm dx^\rho$, which have different dimensions. To observe the difference more clearly, we consider their relationships with the electromagnetic field tensor $F_{\mu\nu}(q)$. The electromagnetic field tensor $F(q)$ in space-time is defined as
}
\begin{equation}
\begin{split}
F(q)&=\frac{1}{2}F_{\rho\sigma}(q)\,\mathrm dx^\rho\wedge\mathrm dx^\sigma\\
&=\frac{1}{2}\left(\partial_\rho A_\sigma-\partial_\sigma A_\rho\right)\,\mathrm dx^\rho\wedge\mathrm dx^\sigma\\
&=\mathrm dA.
\end{split}
\end{equation}
Since $F_{\rho\sigma}=2\partial_{[\rho}A_{\sigma]}=2\partial_{[\rho}\mathcal A_{\sigma]}$,  $F(q)$ can be promoted to the bundle tensor $\mathcal F(q,e)$:
\begin{equation}
\begin{split}
\mathcal F(q,e)&=\frac{1}{2}\mathcal F_{\rho\sigma}(q,e)\,\mathrm dx^\rho\wedge\mathrm dx^\sigma\\
&=\frac{1}{2}\left(\partial_\rho\mathcal A_\sigma-\partial_\sigma\mathcal A_\rho\right)\,\mathrm dx^\rho\wedge\mathrm dx^\sigma.
\end{split}
\end{equation}
Then, $\mathcal F_{\rho\sigma}(q,e)=F_{\rho\sigma}(q)$, that is, $\mathcal F$ is independent of the selection of the tetrad. However,
\begin{equation}
\begin{split}
&\mathrm d\mathcal A=\mathrm d\left(\mathcal A_\rho(q,e)\,\mathrm dx^\rho\right)=\mathrm d\mathcal A_\rho\wedge\mathrm dx^\rho\\
&=\frac{1}{2}\left(\partial_\rho\mathcal A_\sigma-\partial_\sigma\mathcal A_\rho\right)\,\mathrm dx^\rho\wedge\mathrm dx^\sigma+\frac{\partial\mathcal A_\rho}{\partial{\Lambda^\mu}_\nu}\,\mathrm d{\Lambda^\mu}_\nu\wedge\mathrm dx^\rho\\
&=\mathcal F(q,e)+\left(\frac{\partial}{\partial{\Lambda^\mu}_\nu}\frac{\partial}{\partial x^\rho}\Omega\right)\,\mathrm d{\Lambda^\mu}_\nu\wedge\mathrm dx^\rho
\end{split}
\end{equation}
We observe that $\mathcal F\not=\mathrm d\mathcal A$ and $\mathcal F=\mathrm D\mathcal A$, where $\mathrm D$ is the exterior covariant derivative in the tetrad bundle, whose definition can be found in the literature \cite{kobayashi1963foundations}.

It is necessary to discuss the performance of the {projection of $\mathcal A$} to space-time. Only this projection can be directly observed. If we choose a fixed {tetrad} $e(q)$ at each space-time point $q$, {mapping $e(q)$ is called as a tetrad field}. 

Applying the pullback mapping $e^*$ {that is} induced by the tetrad $e(q)$ to $\mathcal A$,
we {obtain} a vector field $e^*\mathcal A$ in space-time. {Via} Eq. (\ref{bundleform}), we {obtain} the component form of $e^*\mathcal A$ {by} setting $e_\nu(q)={\Lambda^\mu}_\nu(q)\partial_\mu$, where ${\Lambda^\mu}_\nu(q)$ is related to $q$,
\begin{equation}
	\begin{split}
		(e^*\mathcal A)_\rho(q)&=\mathcal A_\rho(q,e(q))+{\mathcal A^\nu}_\mu(q,e(q))\frac{\partial{\Lambda^\mu}_\nu(q)}{\partial x^\rho}
		\\&=\mathcal A_\rho(q,e(q));
	\end{split}
\end{equation}
{that is,}
\begin{equation}\label{pro_jisuan}
	\begin{split}
		(e^*\mathcal A)(q)&=\mathcal A_\rho(q,e(q))\,\mathrm dx^\rho\\
		&=\big(A_\rho(x)+\partial_\rho\Omega(x,\Lambda)\big)\,\mathrm dx^\rho,
		\end{split}
\end{equation}
where $\mathrm dx^\rho$ becomes a vector in space-time again.

{Pullback} $e^*$ is a projection. The tetrad $e(q)$ represents the (local) laboratory reference frame that we are observing. We regard $e(q)$ as a Lorentz gauge.

{When} a set of coordinates $x(q)$ is selected, a natural selection of a tetrad is $e(q)=\left(\partial_0,\partial_1,\partial_2,\partial_3\right)$. This is the default choice of optical experiments so that the influence of the local reference frame (tetrad) {can be} ignored. The coordinate of $(q,e(q))$ is $(x^\sigma,{\delta^\mu}_\nu)$. {Since} $\partial_\rho\Omega(x^\sigma,{\delta^\mu}_\nu)=0$, $\left(e^*\mathcal A\right)_\rho(q)=A_\rho(x)$.  If we choose another coordinate $\bar x=\Lambda x$, the  tetrad field will be reselected as $\bar e(q)=\left(\bar\partial_0,\bar\partial_1,\bar\partial_2,\bar\partial_3\right)$. Naturally, $\mathcal A$'s projection under $\bar e(q)$ is
\begin{equation}\label{eA1}
	\left(\bar e^*\mathcal A\right)(q)=\bar{\mathcal A}_\rho(q,\bar e(q))\,\mathrm d\bar x^\rho=\bar A_\rho(\bar x)\,\mathrm d\bar x^\rho.
\end{equation}

From {an} other perspective, to calculate $\bar e^*\mathcal A$, {we observe} that $\bar e(q)=e(q)\Lambda^{-1}$. Substituting $x=\Lambda^{-1}\bar x$ and $\bar e(q)$ into Eq. (\ref{pro_jisuan}), we {obtain}
\begin{equation}\label{eA2}
\begin{split}
		\left(\bar e^*\mathcal A\right)(q)&=\mathcal A_\rho(q,\bar e(q))\,\mathrm dx^\rho
		\\
		&=\big(A_\rho(x)+\partial_\rho\Omega(x,\Lambda^{-1})\big)\,\mathrm dx^\rho
		\\
		&=\big(A_\sigma(\Lambda^{-1}\bar x)+\left(\partial_\sigma\Omega\right)(\Lambda^{-1}\bar x,\Lambda^{-1})\big){\Lambda_\rho}^\sigma\mathrm d\bar x^\rho
\end{split}
\end{equation}
{Contrasting} Eqs. (\ref{eA1}) and (\ref{eA2}), there must be
\begin{equation}\label{pro_trans}
	\bar A_\rho(\bar x)={\Lambda_\rho}^\sigma A_\sigma(\Lambda^{-1}\bar x)+{\Lambda_\rho}^\sigma\left(\partial_\sigma\Omega\right)(\Lambda^{-1}\bar x,\Lambda^{-1}),
\end{equation}
where both $A_\rho(x)$ and $\bar A_\rho(\bar x)$ satisfy the Coulomb gauge, (\ref{kulun}); {hence,} they are not Lorentz vectors in space-time. This is expected {because} the electromagnetic $4$-potential {that} we have observed is a projection of the vector in {the tetrad bundle;} thus, it is {one-sided} and, naturally, cannot satisfy Lorentz covariance. {Vector $\mathcal A(q,e)$ provides a complete description} of the electromagnetic field. Equation (\ref{pro_trans}) proves that the definition of $\mathcal A$ is independent of {the} reference frames. {For any} reference frame $x$, $A_\rho(x)$ in the frame that satisfies the Coulomb gauge is promoted to the vector $\mathcal A$. {Hence,} $\mathcal A$ is the geometric invariant {for which} we are looking.

Since $A_\rho(x)$ in Eq. (\ref{Aqe}) must satisfy the Coulomb gauge, (\ref{kulun}), $\mathcal A$ {does not} change when $\mathscr A_\rho(x)$ {is subjected to} gauge transformation $\mathscr A_\rho(x)+\partial_\rho\epsilon(x)$. {Thus,} $\mathcal A$ has $U(1)$ gauge invariance.

It is natural to replace $\mathscr A_\rho$ and $A_\rho$ with $\mathcal A_\rho$ in {the definition of} each physical quantity. {The form of the electromagnetic field tensor will be unchanged under the replacement since $\mathcal F_{\rho\sigma}=F_{\rho\sigma}$ and $e^*\mathcal F=F$.} {In addition}, the form of the Lagrangian will be unchanged, {in particular, for Maxwell's equation.} However, this substitution {results in} the promotion of scalars, vectors and tensors from space-time to the {tetrad bundle}. 

The definition of the canonical energy-momentum tensor $T_N$ is modified to
\begin{equation}
	\begin{split}
		\mathcal T_N^{\mu\nu}(q,e)=-\frac{1}{4}\eta^{\mu\nu}F_{\rho\sigma}(q)F^{\rho\sigma}(q)+F^{\mu\rho}(q)\partial^\nu\mathcal A_\rho(q,e).
	\end{split}
\end{equation}
Then we can prove that $\mathcal T_N$ is Lorentz covariant:
\begin{equation}
	U(\Lambda)\mathcal T_N^{\mu\nu}(x,\mathbbm 1)U^{-1}(\Lambda)={\Lambda_\rho}^\mu{\Lambda_\sigma}^\nu \mathcal T_N^{\rho\sigma}(\Lambda x,\Lambda).
\end{equation}
{From} the $U(1)$ gauge invariance of $\mathcal A$, the gauge invariance of $\mathcal T_N$ {follows}.

Similarly, the spin $S^{\rho\mu\nu}$ is modified to
\begin{equation}
	\mathcal S^{\rho\mu\nu}=F^{\rho\mu} \mathcal A^\nu-F^{\rho\nu} \mathcal A^\mu.
\end{equation}
The Belinfante energy-momentum tensor $T_B$ remains unchanged under $\mathscr A\to\mathcal A$.

In the laboratory reference frame, we select the tetrad $e(q)=\left(\partial_0,\partial_1,\partial_2,\partial_3\right)$ and the projection of the canonical energy-momentum tensor $\mathcal T_N$ is
\begin{equation}
\begin{split}
	\left(e^*\mathcal T_N\right)^{\mu\nu}(x)
	&=T_N^{\mu\nu}(x)
	\\&=-\frac{1}{4}\eta^{\mu\nu}F_{\rho\sigma}F^{\rho\sigma}+F^{\mu\rho}\partial^\nu A_\rho.
\end{split}
\end{equation}

Naturally, the projection of spin {$\mathcal S$} is
\begin{equation}
	\left(e^*\mathcal S\right)^{\rho\mu\nu}=F^{\rho\mu}A^{\nu}-F^{\rho\nu}A^{\mu}.
\end{equation}

Similarly, we can  {formulate} the projection of {AM} density, which is {the} same as Eq.  (\ref{gauge-invLS}). These results are consistent with eqs.(\ref{gau-invTN}) {and} (\ref{gau-invS}). Equations (\ref{gau-invTN}) and (\ref{gau-invS}) may have Lorentz covariance; however, $T_N$ and $S$, which only reflect part of the information, do not provide a complete physical description, in contrast to $\mathcal T_N$ and $\mathcal S$.

Since the projection of $\mathcal A$ maintains the Coulomb gauge in any reference frame, after being promoted to {bundle tensors}, the conserved orbital and spin {AMs that are} constructed by Bliokh in  \cite{bliokh2014conservation} will be meaningful in any reference frame.

\section{Summary}\label{5}
The problem {that the} orbital and spin {AMs} of {an} electromagnetic field {cannot be  simultaneously} gauge invariant and Lorentz covariant becomes clear in the framework of quantum field theory. With a Lorentz transformation, the electromagnetic $4$-potential that satisfies the Coulomb gauge will produce a phase, which is denoted as $\Omega$, which depends on this transformation and breaks the {Lorentz-covariant} form of {the} electromagnetic $4$-potential. {After promotion as a vector in the tetrad bundle, electromagnetic $4$-potential} can possess Lorentz covariance, which suggests that the {noncovariance} is due to our negligence of the transformation of the local reference frame. {Equivalently}, the electromagnetic object {at} $(x^\mu,\partial_\nu)$ should be covariant with {the object} at $({\Lambda^\mu}_\sigma x^\sigma,{\Lambda_\nu}^\sigma\partial_\sigma)$; {however,} we are accustomed to comparing the electromagnetic objects at $(x^\mu,\partial_\nu)$ and $({\Lambda^\mu}_\sigma x^\sigma,\partial_\nu)$. The local reference frame also plays a role as an internal degree of freedom. 

{The tetrad bundle is} closely related to the gravitational effects \cite{tanii2014introduction}. {Strictly}, in curved space-time, the local frame $e(q)=\left(\partial_0,\partial_1,\partial_2,\partial_3\right)$ is not orthonormal; hence, {it} is not a tetrad {so that the coordinate and tetrad diverge, which will generate observable differences between the electromagnetic 4-potential and the electromagnetic vector in a strong gravitational field. Therefore, the electromagnetic vector could play a role in the optical observation of astronomical objects.} 

In addition, the electromagnetic $4$-potential {is} a gauge potential in the principal bundle whose structure group is $U(1)$ and a projection of a vector in the {tetrad bundle}. We can view the electromagnetic $4$-potential as a link {that connects} the two bundles. A natural question arises: What {type} of relationship has been established by {the electromagnetic $4$-potential} between electromagnetic and gravitational interactions? We will explore this further.

%\bibliographystyle{apsrev4-1}
%\bibliography{References}

\end{document}